\documentclass{article}

\usepackage{arxiv}

\usepackage[utf8]{inputenc} 
\usepackage[T1]{fontenc}    
\usepackage{lmodern}        
\usepackage{hyperref}       
\usepackage{url}            
\usepackage{booktabs}       
\usepackage{amsfonts}       
\usepackage{nicefrac}       
\usepackage{microtype}      
\usepackage{graphicx}

\title{Evaluating undercounts in epidemics: response to Maruotti
\emph{et al.} 2022}

\author{
    Michael Li
   \\
    Public Health Agency of Canada \\
   \\
  \texttt{} \\
   \And
    Jonathan Dushoff
   \\
    McMaster University \\
   \\
  \texttt{} \\
   \And
    David J. D. Earn
   \\
    McMaster University \\
   \\
  \texttt{} \\
   \And
    Benjamin M. Bolker
   \\
    McMaster University \\
   \\
  \texttt{} \\
  }

\newlength{\cslhangindent}
\newlength{\csllabelwidth}
\newlength{\cslentryspacingunit} 
\newenvironment{cslreferences}%
  {}%
  {\par}
 {
  \setlength{\parindent}{0pt}
  \ifodd #1
  \let\oldpar\par
  \def\par{\hangindent=\cslhangindent\oldpar}
  \fi
  \setlength{\parskip}{#2\cslentryspacingunit}
 }%
 {}
\usepackage{calc}

\usepackage[utf8]{inputenc}
\usepackage{grffile}
\begin{document}
\maketitle

\begin{abstract}

\end{abstract}

\hypertarget{abstract}{%
\section{Abstract}\label{abstract}}

Maruotti \emph{et al.} 2022 used a mark-recapture approach to estimate
bounds on the true number of monkeypox infections in various countries.
These approaches are fundamentally flawed; it is impossible to estimate
undercounting based solely on a single stream of reported cases.
Simulations based on a Richards curve for cumulative incidence show
that, for reasonable epidemic parameters, the proposed methods estimate
bounds on the ascertainment ratio of \(\approx 0.2-0.5\) roughly
\emph{independently} of the true ascertainment ratio. These methods
should not be used.

\hypertarget{introduction}{%
\section{Introduction}\label{introduction}}

Several papers\textsuperscript{1--3} have promoted formulas that claim
to provide bounds on the completeness of sampling of infectious disease
cases, based only on case reports. We believe these approaches are
fundamentally flawed, and that it is impossible to estimate
undercounting from incidence data without a specialized sampling design
or some kind of auxiliary information.

The authors use mark-recapture formulas developed by
Chao\textsuperscript{4} and others\textsuperscript{5} to estimate bounds
on true population sizes based on the numbers of individuals observed
multiple times. For example, the proposed estimator for the lower bound
on unobserved individuals (hidden cases) is
\(\Delta N(t) (\Delta N(t) - 1)/(1 + \Delta N(t-1))\), where
\(\Delta N(t)\) is the number of new cases observed per reporting
period; extended formulas adjust for mortality and recovery. The upper
bound also involves \(\Delta N(t-2)\).\textsuperscript{1,3}

\hypertarget{critique}{%
\section{Critique}\label{critique}}

\hypertarget{logical-argument}{%
\subsection{Logical argument}\label{logical-argument}}

This approach misuses the mark-recapture formulas. Cases identified at
time \(t-1\) are claimed to be representative of the number of cases
counted twice: why? The fact that the same individual \emph{could} be
counted twice in the cumulative case report (for some sampling designs)
is irrelevant. How can comparing yesterday's count to today's provide
information about the completeness of sampling?

In principle, the number of unobserved hidden cases can be estimated if
cases can be re-identified, or even with unmarked/unidentified cases
given an appropriate sampling design.\textsuperscript{6} In practice
public health case reporting rarely uses such sampling designs. Case
reporting is usually exclusive (i.e.~someone who has been identified as
a case will not be reported again later), or anonymized so that we
cannot identify a particular infected individual as double-counted.
Mark-recapture methods can provide valuable public health information,
but ``one needs at least two sources of information with individual case
reporting and a unique personal identifier for each
case''.\textsuperscript{7}

\hypertarget{simulation-example}{%
\subsection{Simulation example}\label{simulation-example}}

We ran simulations using a Richards curve for the cumulative incidence
of the epidemic.\textsuperscript{8} We computed expected incidence by
differencing the cumulative incidence, drew a random negative binomial
deviate with mean equal to the expected incidence, and used a binomial
sample with probability equal to the ascertainment ratio \(a\) to get
the number of observed cases. Throughout, we used a shape parameter of
\(s=2\) and a final epidemic size of \(10^5\) for the Richards curve,
and a negative binomial dispersion parameter \(k=5\). We varied the
reporting period (\(\Delta t = \{1, 7\}\)); starting incidence
(\(I_0 = \{20, 40\}\)); epidemic growth rate (\(r\) = 0.01 to 0.08 per
day); and ascertainment ratio (\(a\) from 0.05 to 0.6). We ran each
simulation for 100 days and used the R package
\texttt{asymptor}\textsuperscript{9} to compute bounds on the
ascertainment ratio.

The authors indicated (pers. comm.) that they intended the estimator to
be used at the beginning of an epidemic. Therefore we considered only
sample points when the number of cases was between 5 and 500 (exclusive)
and the lower bound estimator for hidden cases was greater than 1.

For each simulation run (80 in total), we computed the mean and
confidence intervals for the estimated lower and upper bounds of
\(\hat a\) over time (Figure 1). The bounds on \(\hat a\) rarely overlap
the true value, and are \emph{largely independent of the true values of
\(a\)}. The only noticeable signal arises from the bias-correction
terms: simulations with lower overall case numbers (low \(r\), low
\(a\), \(\Delta t = 1\)) have larger lower bounds and smaller upper
bounds. In simulations without noise and with the simpler,
non-bias-corrected expression for the lower bound (not shown), the
lower-bound estimates of \(\hat a\) are completely independent of \(a\);
some algebra shows that during the exponential growth phase of an
epidemic, the (simplified) lower bound on \(\hat a\) is exactly equal to
\(1/(1+\exp(r \Delta t))\).

\begin{figure}
\centering
\includegraphics{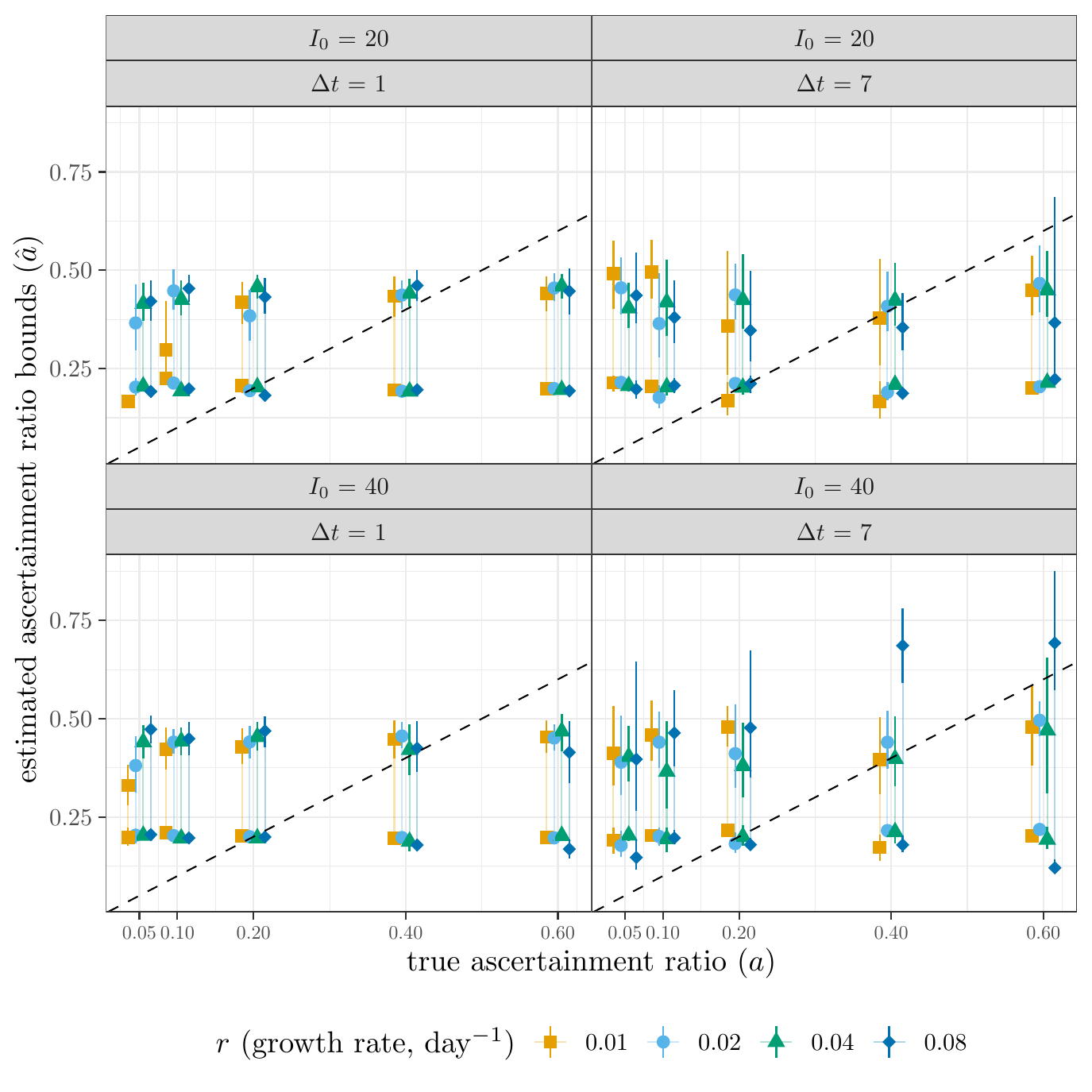}
\caption{Comparison of true ascertainment ratio (\(a\)) to estimated
lower and upper bounds of ascertainment ratio (\(\hat a\)). Dashed line
is the one-to-one line (estimated = true).}
\end{figure}

We conclude that the authors' formulas appear to work well because they
lead to plausible bounds on the ascertainment ratio (\(\approx\) 0.2 --
0.5) for realistic values of the epidemic growth rate, but that they are
in fact nearly unrelated to the true ascertainment ratio and should not
be applied to disease outbreak incidence data.

\begin{center}\rule{0.5\linewidth}{0.5pt}\end{center}

Further details, and source code for all examples, are available at
\url{https://github.com/wzmli/undercount/}.

\hypertarget{references}{%
\subsection*{References}\label{references}}
\addcontentsline{toc}{subsection}{References}

\hypertarget{refs}{}
\begin{cslreferences}
\leavevmode\hypertarget{ref-bohningEstimating2020}{}%
1. Böhning, D., Rocchetti, I., Maruotti, A. \& Holling, H. Estimating
the undetected infections in the Covid-19 outbreak by harnessing
capture--recapture methods. \emph{International Journal of Infectious
Diseases} \textbf{97}, 197--201 (2020).

\leavevmode\hypertarget{ref-maruottiEstimating2022}{}%
2. Maruotti, A., Böhning, D., Rocchetti, I. \& Ciccozzi, M. Estimating
the undetected infections in the Monkeypox outbreak. \emph{Journal of
Medical Virology} 1--4 (2022)
doi:\href{https://doi.org/10.1002/jmv.28099}{10.1002/jmv.28099}.

\leavevmode\hypertarget{ref-rocchettiEstimating2020}{}%
3. Rocchetti, I., Böhning, D., Holling, H. \& Maruotti, A. Estimating
the size of undetected cases of the COVID-19 outbreak in Europe: An
upper bound estimator. \emph{Epidemiologic Methods} \textbf{9}, (2020).

\leavevmode\hypertarget{ref-chaoEstimating1989a}{}%
4. Chao, A. Estimating Population Size for Sparse Data in
Capture-Recapture Experiments. \emph{Biometrics} \textbf{45}, 427
(1989).

\leavevmode\hypertarget{ref-alfoUpper2021}{}%
5. Alfò, M., Böhning, D. \& Rocchetti, I. Upper bound estimators of the
population size based on ordinal models for capture‐recapture
experiments. \emph{Biometrics} \textbf{77}, 237--248 (2021).

\leavevmode\hypertarget{ref-royleHierarchical2008}{}%
6. Royle, J. A. \& Dorazio, R. M. \emph{Hierarchical modeling and
inference in ecology: The analysis of data from populations,
metapopulations and communities}. (Academic Press, 2008).

\leavevmode\hypertarget{ref-desenclosLimitations1994}{}%
7. Desenclos, J.-C. \& Hubert, B. Limitations to the Universal use of
Capture-Recapture Methods. \emph{International Journal of Epidemiology}
\textbf{23}, 1322--1323 (1994).

\leavevmode\hypertarget{ref-maEstimating2014}{}%
8. Ma, J., Dushoff, J., Bolker, B. M. \& Earn, D. J. D. Estimating
Initial Epidemic Growth Rates. \emph{Bulletin of Mathematical Biology}
\textbf{76}, 245--260 (2014).

\leavevmode\hypertarget{ref-grusonAsymptor2020}{}%
9. Gruson, H. asymptor: Estimate the lower and upper bound of
asymptomatic cases in an epidemic using the capture/recapture methods
(package version 1.0). (2020).
\end{cslreferences}

\bibliographystyle{unsrt}
\bibliography{undercount.bib}

\end{document}